\documentclass[9pt,twocolumn,twoside]{opticajnl}

\journal{ol} 

\setboolean{shortarticle}{true}

\usepackage{lineno}

\title{Microfluidic Filling and Spectroscopy of Colloidal CdSe/CdS Nanoplatelets in Liquid Core Fibers}

\author[1,*]{Simon Spelthann}
\author[1]{Dan Huy Chau}
\author[2,3]{Lars F. Klepzig}
\author[2,3]{Dominik A. Rudolph}
\author[4]{Mario Chemnitz}
\author[4]{Saher Junaid}
\author[1,3,5,6]{Detlev Ristau}
\author[4,7]{Markus A. Schmidt}
\author[2,3,6,8]{Jannika Lauth}
\author[1,3,5]{Michael Steinke}

\affil[1]{Leibniz University Hannover, Institute of Quantum Optics, Welfengarten 1, D-30167 Hannover, Germany}
\affil[2]{Leibniz University Hannover, Institute of Physical Chemistry and Electrochemistry, Callinstraße 3A, D-30167 Hannover, Germany}
\affil[3]{Cluster of Excellence PhoenixD (Photonics, Optics, and Engineering – Innovation Across Disciplines), Welfengarten 1A, D-30167 Hannover, Germany}
\affil[4]{Leibniz Institute of Photonic Technology, Albert-Einstein-Straße 9, D-07745 Jena, Germany}
\affil[5]{Leibniz University Hannover, QUEST-Leibniz-Research School, Callinstraße 36, D-30167 Hannover, Germany}
\affil[6]{Leibniz University Hannover, Laboratory of Nano and Quantum Engineering (LNQE), Schneiderberg 39, D-30167 Hannover, Germany}
\affil[7]{Otto Schott Institute of Material Research, Fraunhoferstraße 6, D-07745 Jena, Germany}
\affil[8]{University Tübingen, Institute of Physical and Theoretical Chemistry, Auf der Morgenstelle 18, D-72076 Tübingen, Germany}

\affil[*]{Corresponding author: spelthann@iqo.uni-hannover.de}

\begin{abstract}
Colloidal 2D semiconductor nanoplatelets are highly efficient light emitters, which exhibit large absorption and emission cross sections, and constitute promising laser gain media. 
However, if dispersed in solutions, such nanoplatelets lack a suitable optical platform for scalable and application-oriented integration into optical setups such as lasers.
Here, we demonstrate the first successful integration of solution-processed 2D CdSe/CdS Core/Crown nanoplatelets in m-scale liquid core optical fibers. 
We compare the nanoplatelets' spectroscopic properties before and after filling them into the fibers and find that spontaneous emission is shifted and broadened. We even observe a first evidence of stimulated emission at high excitation energies.
In conclusion, liquid core fibers constitute a novel and scalable platform for optical integration of nanoplatelets for applications as novel, highly reconfigurable laser gain medium.
\end{abstract}

\begin{document}

\maketitle

\section{Introduction}
Two-dimensional (2D) semiconductor nanoplatelets (NPLs) constitute a prominent class of nanocrystals. They exhibit large absorption and emission cross-sections as high as a few 10$^{-12}$\,cm$^{2}$ \cite{Yeltik2015, Taghipour2020}, low Auger recombination rates \cite{Taghipour2020,Kunneman2013,Altintas2019}, and narrow emission linewidths \cite{Ithurria2011,Geiregat2019NPJ}. These properties make NPLs interesting gain media for lasers with unique properties such as low lasing thresholds \cite{Geiregat2018,Geuchies2022}, tunable visible emission, and gain cross sections as high as 10$^{-14}$\,cm$^{2}$ \cite{Guzelturk2014,Tomar2019,Geuchies2020,Delikanli2021CL}. Consequently, in-depth work was done to study the NPL gain, which relies on bound electron-hole states (excitons) \cite{Pietryga2016,Geiregat2019}, and to make it useable for applications. For example, in 2015, Guzelturk et al.\ reported amplified spontaneous emission (ASE) at around 640\,nm with a sub J cm$^{-2}$ threshold from a vertical cavity surface-emitting laser (VCSEL) based on spin-coated CdSe/CdS core/shell NPLs \cite{Guzelturk2015}. Other approaches employed solution-processed colloidal NPLs with benefits that range from flexibility in the cavity design to a high photostability enabled by an easy-to-perform NPL replacement \cite{Tan2019}. For example, Kazes et al.\ \cite{Kazes2002} and Ye et al.\ \cite{Yu2020} demonstrated transversal high-Q whispering gallery mode operation of SiO$_2$ tubes filled with solution-processed NPLs. Using a microfluidic waveguide, Maskoun et al.\ combined solution-processed NPLs with microfluidic waveguides of 4\,mm length, whose endfacets were coated with silver mirrors \cite{Maskoun2021}. Yet other approaches consisted of grated waveguides filled with NPL dispersions \cite{Gheshlaghi2021} or short (mm-scale) pieces of un-coated and non-sealed silica capillaries as longitudinal waveguides \cite{Delikanli2022, Delikanli2021CL}. These results demonstrate that NPLs are excellent gain media. However, the utilized optical platforms lack scalability and flexibility. Such properties are crucial since they drive the transition of NPLs toward applications.\\
Here, we establish a scalable platform for NPLs that addresses the limitations of former approaches by using application-proven fused silica liquid core fibers (LCFs). Such LCFs and similar types of holey microstructured fibers emerged in recent years as a unique platform, e.g.\ to utilize nonlinear effects in gas-filled \cite{Debord2019} or liquid-filled  \cite{Schaarschmidt2020,Xue2022,Scheibinger2023} fibers. For highest integration, we manufactured optofluidic mounts, which allowed us to integrate the LCFs in a monolithic and sealed setup. We successfully filled CdSe/CdS core/crown NPLs dispersed in toluene in bendable LCFs of up to 0.5\,m length. This provides a significantly increased interaction length compared to previous approaches. To validate the filling, we performed optical spectroscopy that allowed us to observe not only spontaneous but also stimulated emission from higher-order excitons. Altogether, our fiber-based approach constitutes a flexible and rugged (sealed) platform for highly integrated fiber lasers with NPLs as gain medium.

\section{Synthesis and Spectroscopy of Colloidal CdSe/CdS Core/Crown Nanoplatelets}
We synthesized colloidal CdSe NPLs with a thickness of 4 monolayers (MLs), i.e.\ 4 atomic layers of Se and 5 atomic layers of Cd, by wet chemical methods as described in \cite{Ithurria2011}. According to Schlosser et al., a laterally grown crown of CdS suppresses non-radiative decays via trap states at the NPLs' edges and increases the quantum yield (QY) from around 10\,\% up to 70 -- 90\,\% without modifying the bandgap \cite{Schlosser2020,Tessier2014}. We stabilized the as-synthesized NPLs with oleic acid. Such colloidal dispersions of NPLs can be stable for months but tend to agglomerate when the ligand concentration is changed, e.g.\ due to dilution. Therefore, we ensured the colloidal stability, taking into account suitable ligand concentrations and interactions within the fiber.
\begin{figure}[htb]
\centering
\includegraphics{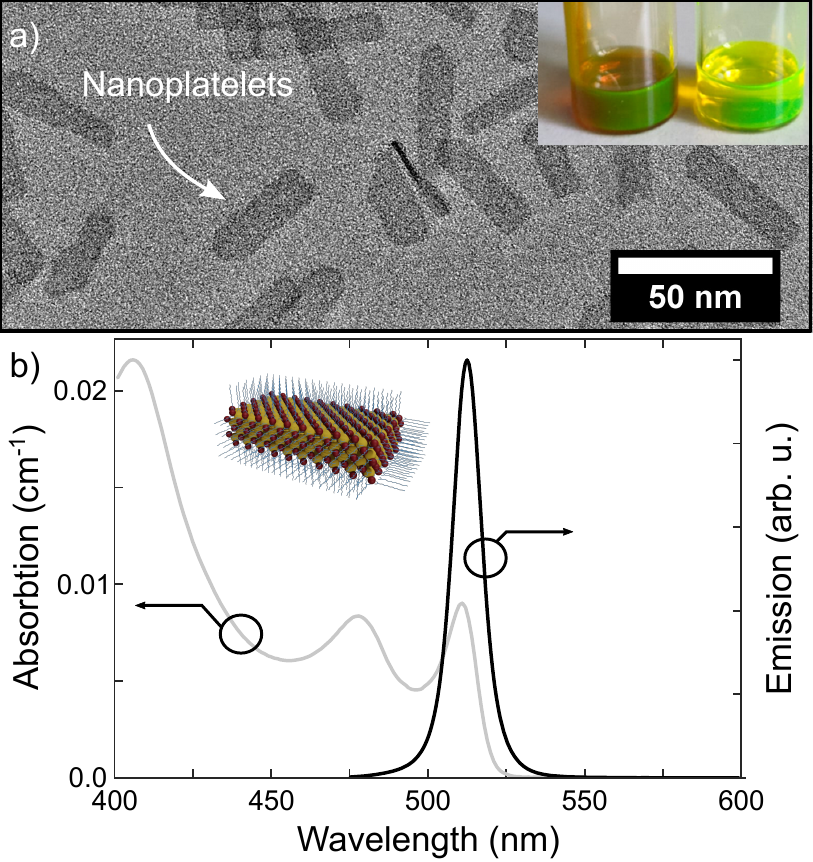}
\caption{Properties of the synthesized CdSe/CdS core/crown NPLs consisting of four monolayers. a) Exemplary transmission electron microscopy picture; inset: optical impression of the dispersed NPLs.
b) Absorbance and normalized emission spectra of the NPLs measured in a 1\,cm cuvette.}
\label{fig:NPLs}
\end{figure}
The NPLs are dispersed in an organic solvent, which fulfills both a chemical and an optical functionality. Chemically, the solvent contains the NPLs and, thus, should form a stable dispersion. Optically, the solvent constitutes the core medium of the fiber and, thus, needs a refractive index higher than that of fused silica. In addition, since the short NPL lifetime requires high-energy pulsed excitation, the to-be-used solvent should have a low nonlinear Raman response. These requirements rule out many organic solvents and led us to use toluene (C$_7$H$_8$, n=1.4968 at 587.6\,nm) in our experiments.\\
As shown in Figure \ref{fig:NPLs}b, the NPLs absorb light over a broad spectral range up to 525\,nm and show spontaneous emission centered at 515\,nm with a FWHM of 10.5\,nm. The narrow optical features are a result of the NPLs' uniform thickness, as the thickness distribution is strongly monodisperse (4 ML). We measured the excited state lifetime and found a monoexponential (single-exciton) decay with a time constant of 4 \,ns. Additionally, we measured the NPLs' quantum yield to be 40\,\%.
\section{Optofluidic Setup and Filling}
\begin{figure}[bth]
\centering
\includegraphics{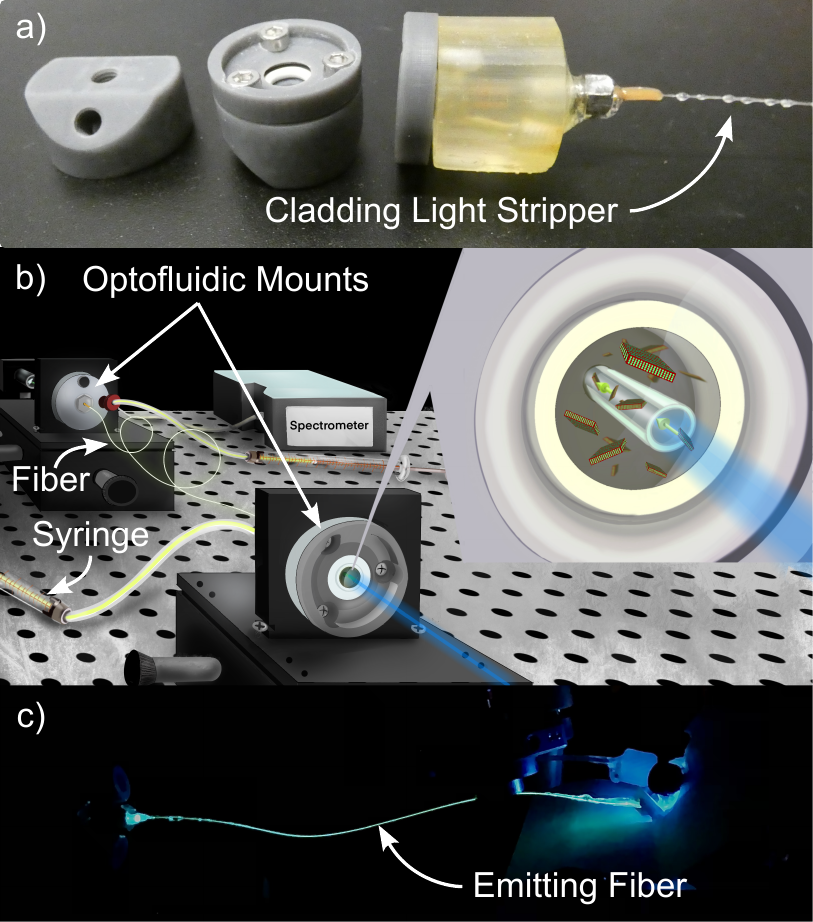}
\caption{Details of the optofluidic setup. a) left: Additively manufactured reservoir of the optofluidic mount; middle: assembled optofluidic mount; right: optofluidic mount with mounted fiber (grey parts: 3D printed; clear parts: fused silica). b) Illustration of the optical setup. c) Emission of the NPLs in the fiber under excitation at 480\,nm.}
\label{fig:Setup}
\end{figure}

The fibers used in this work exhibited outer diameters of 125\,$\mu$m or 200\,$\mu$m with inner capillary diameters of 3\,-\,60\,$\mu$m. We manufactured each capillary fiber by fiber-drawing a fused silica tube to the desired outer diameter at temperatures around 2000\,$^\circ$C. The inner channel diameter was measured by SEM or optical microscopy. All fibers were coated with a UV-cured dual-layer acrylat coating providing the well-known mechanical robustness of optical fibers.\\
To prevent the organic solvent from evaporation, the capillary fiber needs to be sealed gas and liquid tight. In addition, the fiber end facets have to be optically accessible to allow light coupling. We achieved that by employing optofluidic mounts, see Figure \ref{fig:Setup}a. The optofluidic mounts exhibit a reservoir (volume of 90\,$\mu$L), which can be filled with the NPL dispersion and makes the fiber optically accessible through an optical window (fused silica). In contrast to previous studies \cite{Chemnitz2017}, we additively manufactured the optofluidic mounts in a stereolithography 3D printer (Formlabs, Form 3) using a resin (Formlabs, Clear Resin V4) resistant against the organic solvents. This approach allowed us to flexibly adjust the mount design to the experimental conditions. As shown in Figure \ref{fig:Setup}a, optical adhesive (Norland Optical Adhesive 68) was used for sealing and, due to its high refractive index, as a cladding light stripper.
We observed that a contamination of the dispersion by the adhesive must be prevented since it degrades the NPLs' chemical stability.\\
For each experiment, we mounted both ends of around 30 to 40\,cm capillary fiber in the optofluidic mounts and filled it with the NPL dispersions. This provides interaction lengths, which are more than 10 times higher than in previous publications \cite{Maskoun2021, Guzelturk2015, Kazes2002, Yu2020}. Typically, we set the NPL concentration in the dispersion to an absorbance of 1 (corresponding to an absorption of 0.8\,dB/cm in a 30\,cm fiber) to obtain an absorption of around 90\,\% through the fiber.\\
For the actual filling process, we investigated different approaches discussed in the following. The first approach corresponds to injecting the NPL dispersion in the reservoir of one optofluidic mount with a syringe. Afterwards, the fiber is filling itself via capillary forces. The filling time can be estimated by Washburn's equation, see. e.g.\ \cite{Lee11}. Typically, the meniscus was observed visually and once the fiber was completely filled, the reservoir of the second mount was filled too. However, in some cases, the filling stopped in the middle of the procedure and static air bubbles were observable inside the fiber. To remove these bubbles, the syringe was used to apply pressure to the first optofluidic mount, which provided an additional filling force in addition to the inherent capillary forces. 
These approaches yielded a relatively fast filling of the fiber with flow velocities of up to 27\,cm/s within the capillary. In some cases, strong shear forces induced by such high flow velocities must have removed some of the NPLs' ligands as we have observed a degradation of the QY, which is a indicator of NPL clustering induced by ligand removal. To obtain slower NPL flow velocities, we mounted the optofluidic mounts to the fiber and filled one mount with the pure solvent, which filled the fiber via capillary forces. Afterwards, we filled the second mount with the dispersion in order to let the NPLs slowly penetrate the fiber via Brownian motion. However, the success and reproducibility of this approach was limited as well. Therefore, even though the reproducibility of the filling via capillary forces needs to be further increased, we successfully applied it for our experiments.

\section{Spectroscopic Validation of the NPL Filling}
To validate the successfully filling of the NPLs, we performed absorption and emission spectroscopy. For absorption measurements, we coupled pulses of a supercontinuum source (NKT, SuperK Compact) into the filled fiber core. The pulses cover a spectral range from 450\,nm to 2400\,nm with energies of less than 5.5\,$\mu$J at 2\,ns duration (maximum power of 110\,mW at 20\,kHz). We measured the spectra of the transmitted pulses with an USB spectrometer (Ocean Optics, FX-VIS-NIR-ES) as depicted in Figure \ref{fig:Setup}b as a qualitative indicator of the filling success.\\
For emission spectroscopy, we used an optical parametric oscillator (OPO), which is tunable over a broad spectral range from 450\,nm to 2750\,nm and provides pulse energies up to 70\,mJ at 4\,ns pulse duration and 10\,Hz repetition rate. Such high excitation energies and short pulse lengths are necessary to overcome the relatively short NPL excited state lifetime and to obtain population inversion \cite{Spelthann}, which is required for potential stimulated emission. In most experiments, we tuned the OPO to around 480\,nm, well below the bandgap of the 4ML NPLs (510\,nm, c.f. Figure \ref{fig:NPLs}b). Pumping the fiber, we observed a green glow of the fiber (c.f. Figure \ref{fig:Setup}c), which corresponds to the expected green emission. For initial spontaneous emission measurements, the pulse energy was attenuated to a maximum of a few $\mu$J. The emission spectra were recorded with above-mentioned spectrometer. In addition, the NPL emission at a higher pulse energy regime of some mJ was studied.
\begin{figure}[hbt]
\centering
\includegraphics[width=\linewidth]{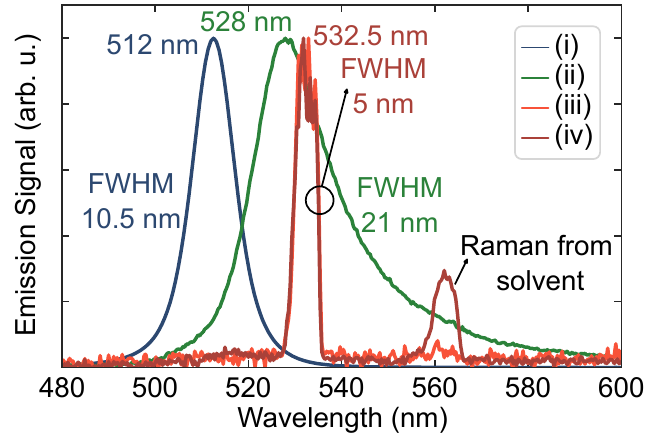}
\caption{Spontaneous emission of the NPLs in a cuvette (i) and in the fiber core (ii). Stimulated emission of the NPLs in the fiber core at pump energies of 1.1\,mJ (iii)  and 2.3\,mJ (iv).}
\label{fig:emissions}
\end{figure}
As a result, Figure \ref{fig:emissions} shows four normalized NPL emission spectra. Spectrum (i) corresponds to the spontaneous emission of the NPLs prior to the filling as presented in Figure \ref{fig:NPLs}b. Spectrum (ii) is the spontaneous emission of the NPLs from inside the fiber at an OPO pump energy of 21\,$\mu$J. We assume that the in-coupling efficiency was rather low (a few percent) and carefully (worst case assumptions) estimated a pump fluence of <\,17\,$\mu$J\,$\mu$cm$^{-2}$ inside the fiber core. An exact determination of the fluence considering further effects such as scattering losses remains a complex task to be addressed in future work. With respect to the spontaneous emission outside the fiber, there is a clear shift of the center wavelength to 528\,nm and an asymmetric broadening with a FWHM of 21\,nm. Such shifting and broadening of the NPLs' spontaneous emission within a confined two-dimensional volume has been observed before \cite{Delikanli2022, Delikanli2021CL}. Here, we attribute it to three effects, which interact and also reduce the QY. First, the electronic environment inside the fiber, in particular due to well-known OH$^{-}$ surface groups \cite{Bise2004}, modifies the NPLs' electronic band structure. Second, emitted light is reabsorbed along the fiber, which is a common effect in optically active fibers \cite{Li13} and is particularly significant in active media with small Stokes shifts such as the NPLs. Third, emission from higher exciton states is red shifted \cite{Taghipour2020}.\\
Finally, spectrum (iii) and (iv) correspond to the emissions at OPO pump energies of 1.1\,mJ and 2.3\,mJ, which yielded estimated in-core pump fluences of <\,360\,$\mu$J\,cm$^{-2}$ and <\,420\,$\mu$J\,cm$^{-2}$. The shifted emission at around 532.5\,nm and the reduced FWHM of 5\,nm indicates the onset of stimulated emission (i.e.\ ASE) along the fiber. Note that such ASE at similar pump fluence levels has been reported for other NPL laser setups, see e.g.\ \cite{Delikanli2022, Delikanli2021CL}. As discussed elsewhere \cite{Taghipour2020}, stimulated emission can only occur via higher-order exciton states, which are associated with a red shift of the emission due to bandgap absorption bleaching. The additional emission at around 562\,nm, which occurs at the highest pump energy, belongs to the onset of nonlinear Raman scattering of toluene (shift of 3000\,cm$^{-1}$) \cite{Wilmshurst1957}. 

\section{Conclusions and Outlook}
We synthesized 4ML CdSe/CdS core-crown NPLs and filled their dispersions in capillary fibers. We ensured colloidal stability by covering the NPLs with surface ligands and developed monolithic optofluidic fiber mounts. To maintain stability during fiber filling, contamination of the dispersions by optical adhesives must be prevented. Furthermore, it is important to carefully seal the mounts to avoid air bubbles. We performed optical emission spectroscopy of the NPLs in the LCF and compared the resulting spectra to the NPLs' emission spectra in a cuvette. At low excitation fluences (<\,17\,$\mu$J\,cm$^{-2}$), the emission in the fiber shifts and broadens compared to the cuvette, which we explain by the modified electronic environment, reabsorption, and higher-order exciton contributions. At high excitation fluences (<\,360\,$\mu$J\,cm$^{-2}$), the emission shifts further and narrows down, which is a indicator for stimulated emission (i.e.\ ASE) via higher-order excitons as discussed elsewhere \cite{Delikanli2022, Delikanli2021CL}.\\
Altogether, capillary fibers, which are microfluidically filled with NPL dispersions constitute a flexible and rugged (sealed) platform for highly integrated fiber lasers that use NPLs as gain medium. Since colloidal chemistry allows to tune the emission wavelength through particle size or shape, our approach will allow to cover a broad range of wavelengths and we conducted first experiments towards 5 ML NPLs. However, other wavelengths come with new challenges, e.g.\ seeds for Raman scattering. In future works, we aim to further develop our approach to investigate not only ASE but also lasing operation (e.g. by utilizing an external cavity) in NPL filled LCF.

\begin{backmatter}
\bmsection{Author Contributions} 
\textbf{Conceptualization:} S.S., J.L, M.S.; 
\textbf{Methodology:} S.S., L.F.K., J.L, M.S.; 
\textbf{Validation:} S.S., D.H.C, L.F.K., D.A.R., J.L., M.S.; 
\textbf{Formal Analysis:} S.S., L.F.K., D.A.R., S.J., M.S.; 
\textbf{Investigation:} S.S., D.H.C, L.F.K., D.A.R., S.J.; 
\textbf{Resources:} J.L., D.R., M.S.; 
\textbf{Data Curation:} S.S., D.H.C., L.F.K., D.Ru., S.J., M.S.; 
\textbf{Writing - Original Draft:} S.S., M.S.; 
\textbf{Writing - Review \& Editing:} S.S., L.F.K., M.C., M.A.S., D.R., J.L., M.S.; 
\textbf{Visualization:} S.S., L.F.K.; 
\textbf{Supervision:} M.C., M.A.S., J.L., M.S.; 
\textbf{Funding Acquisition:} J.L., M.S.
\bmsection{Funding} S.S., D.H.C., L.F.K., D.A.R., D.R., J.L. and M.S. thank the Deutsche Forschungsgemeinschaft (DFG, German Research Foundation) for partly funding this work under Germany's Excellence Strategy within the Cluster of Excellence PhoenixD (EXC-2122, 390833453). The work of S. S. and M. S. was also partly funded by: Deutsche Forschungsgemeinschaft (DFG, German Research Foundation) under Germany's Excellence Strategy – EXC-2123 Quantum Frontiers – 390837967. M.C. acknowledges funding from the Fonds de recherche du Quebec, Nature et technologies (FRQNT) via the PBEEE program and the Carl-Zeiss Stiftung via the NEXUS program. M.A.S. acknowledges support from the Deutsche Forschungsgemeinschaft (DFG, German Research Foundation) withthin the research grants SCHM2655/3-2 and JU 3230/1-1.

\bmsection{Acknowledgments} The authors thank K. Hausmann and R. Stephan for drawing the capillary fibers, B. Kraus for providing access to the stereolithography printer, A. Niebur for taking TEM pictures, and N. Bigall and D. Dorfs for access to their spectrometers. The Laboratory for Nano and Quantum Engineering (LNQE) provided access to the transmission electron microscope (TEM).

\bmsection{Disclosures} The authors declare no conflicts of interest.

\bmsection{Data availability} Data underlying the results presented in this paper are not publicly available at this time but may be obtained from the authors upon reasonable request.

\end{backmatter}





\bibliography{sample}

\bibliographyfullrefs{sample}

\end{document}